# Effective medium approximation of ellipsometric response from random surface roughness simulated by finite-element method

B. Fodor[1,2,*], P. Kozma[1], S. Burger[3], M. Fried[1,4], P. Petrik[1,4]

[1]*Institute for Technical Physics and Materials Science (MFA), Centre for Energy Research of the Hungarian Academy of Sciences, Konkoly Thege út. 29-33, H-1121 Budapest, Hungary*

[2]*Doctoral School of Molecular- and Nanotechnologies, Faculty of Science, University of Pécs, H-7624 Pécs, Ifjúság útja 6, Hungary*

[3]*Zuse Institute Berlin (ZIB), Takustrasse 7, D-14195 Berlin, Germany*

[4]*Doctoral School of Molecular- and Nanotechnologies, Faculty of Information Technology, University of Pannonia, Egyetem u. 10, Veszprém, H-8200, Hungary*

*Corresponding author. Electronic address: fodor@mfa.kfki.hu    +3620 361 7876

We used numerical simulations based on the finite element method (FEM) to calculate both the amplitude and phase information of the scattered electric field from random rough surfaces, which can be directly compared to ellipsometric measurements and effective medium approximation (EMA) calculations. FEM can serve as an exploration tool for the relationship between the thickness of the surface roughness evaluated by Bruggeman EMA and the morphological parameters of the surface, such as the root mean square height, the lateral auto-correlation length and the typical average slope. These investigations are of high interest in case of poly-crystalline and amorphous materials. The paper focuses on the simulations of rough Si surfaces. The ellipsometric calculations from FEM and EMA simulations match for wavelengths of illumination much shorter than the typical feature size of the surface. Furthermore, for these cases, the correlation between the EMA thickness and the root mean square height of the roughness for a given auto-correlation length is quadratic, rather than linear, which is in good agreement with experimental measurements and analytical calculations presented in recent reports.

## 1. Introduction

Characterizing surface roughness with ellipsometry has become a routine practice since the birth of spectroscopic ellipsometry because of its fast, non-destructive and in-line capabilities. The most widely used models describe the surface roughness with an effective medium approximation (EMA), i.e. the surface roughness is considered a homogeneous layer with an effective dielectric function mixed from the dielectric functions of the two media separating the rough interface. A good review about the relationship between surface morphology and EMA roughness can be found in Ref. [1]. Many experimental comparisons have been made between the EMA measured by ellipsometry and the morphology measured by atomic-force microscopy (AFM) for different Si samples: wet etched and thermally annealed Si [2], CVD deposited poly-Si [3, 4] and poly-Si-on-oxide [5], as well as for *in-situ* growth of amorphous hydrogenated Si [6] and CVD deposited microcrystalline-Si on amorphous Si [7]. These works all concluded at a positive linear relationship between EMA roughness and AFM root mean square height, but all with different linear parameter values (slope and offset). One study even showed a negative correlation [8], stating that AFM measurements indicate an increase in root mean square height while ellipsometry suggests a smoothening of roughness. To better grasp the kaleidoscope of these different results, the present study simulates the ellipsometric response of a large number of random Si surfaces with well-defined root mean square heights and correlation lengths. The numerical simulations have been made by finite-element methods (FEM). FEM is a numerical technique to find approximate solutions of partial differential equations. Optical FEM is based directly on the linear Maxwell's equations in frequency domain. Computation of the electric (and the magnetic) field amplitudes are solved on a polygonal mesh, typically triangular, with piecewise-polynomial interpolation between the mesh points. Arbitrary geometrical objects can be defined with permittivity and permeability values assigned to each object (more specifically, assigned to the mesh points approximating the object). A summary of the vast areas of interest of the optical FEM can be found in Ref. [9]. The ellipsometric simulations of the random rough surfaces may be considered in our case as the "measured" samples and the effective medium roughness as the model to be fitted. This approach reveals many interesting effects concerning the relationship between the surface morphology and the thickness of the EMA roughness.

## 2. Model structures

Electromagnetic near fields resulting from plane wave illumination of silicon surfaces with roughness were simulated using the finite-element solver JCMSuite (version 2.16). Specular reflection amplitudes (and intensities) were obtained from far field results computed in post-process as a spatial (discrete) Fourier spectrum. Although the Maxwell equations are solved as stationary wave solutions in frequency domain, from the complex scattered electric fields both the amplitude and phase information can be obtained. As the electric fields of the incident plane waves polarized parallel (*P*) and perpendicular (*S*) to the plane of incidence in the finite-element simulations are defined with unit amplitudes, the ellipsometric complex $\rho$ is obtained as the ratio of the reflected complex amplitudes of the *P* and *S* polarizations. The ellipsometric angles are defined in the usual way as $\Psi = tan^{-1}(\rho)$ and $\Delta = arg(\rho)$, where *tan Ψ* is the amplitude ratio and *Δ* is the phase difference, respectively, of the complex reflection coefficients of *P* and *S* polarized light [10, 11]. The spectra were simulated in a wavelength range from 200 to 1000 nm, in steps of 10 nm, for the angles of incidence of 65° and 75°. The near-field amplitudes had to be computed individually for each wave vector of the illumination, because of the optical dispersion of the Si material [12].

For computational reduction, the simulation domain was 2-dimensional, with a translational symmetry in the direction perpendicular to the plane of incidence. This very useful simplification is based on the assumption that cross-polarizations due to the anisotropic nature of the simulated surface (as opposed to a real randomly rough 2D surface) are negligible, as the surface features are much smaller than the wavelength of illumination (*λ*). Furthermore, to eliminate scattering-like artifacts at the edge of the surface, periodic boundary conditions were used at these lateral sides of the computational domain. For the two remaining sides, transparent boundary condition was applied. The topographic points of the surface were generated with D. Bergström's Open Source MATLAB code [13] in such a way that the height distribution followed a Gaussian statistics. For visualization, a portion of the simulation mesh of a surface with a correlation length of 10 nm and a root mean square roughness of 2.5 nm is shown in Fig.1a (left) with the height distribution histogram (right). An easy way to achieve such a height distribution is to convolute a predefined Gaussian filter on an uncorrelated (Gaussian) distribution of surface points generated by random numbers (i.e. white noise) [14]. The advantage of this approach is that the standard deviation of the uncorrelated distribution and of the Gaussian filter will be inherited, and account for the root mean square height ($R_{RMS}$) and the correlation length ($\xi$) of the surface, respectively. Of course, due to the stochastic nature of the structure, small deviations will be present between the predefined standard deviations and the $R_{RMS}$ values. To achieve adequate Gaussian statistics and diminish deviations from nominal values, the length of the surface to be simulated (*L*) was chosen such that $L/\xi \geq 500$. Additionally, *L* was at least 5 μm so that diffraction due to periodic boundary conditions would be negligible (parameter convergences as a function of *L* were studied). The simulated topographical parameters for $\xi$ were 2.5, 5, 10 and 20 nm, while for the $R_{RMS}$ were 0.5, 1, 1.5, 2.5, 3.5, 5, 7.5, 10, 15 and 20 nm. The combinations of all these parameter values are simulated, totaling in 40 points.

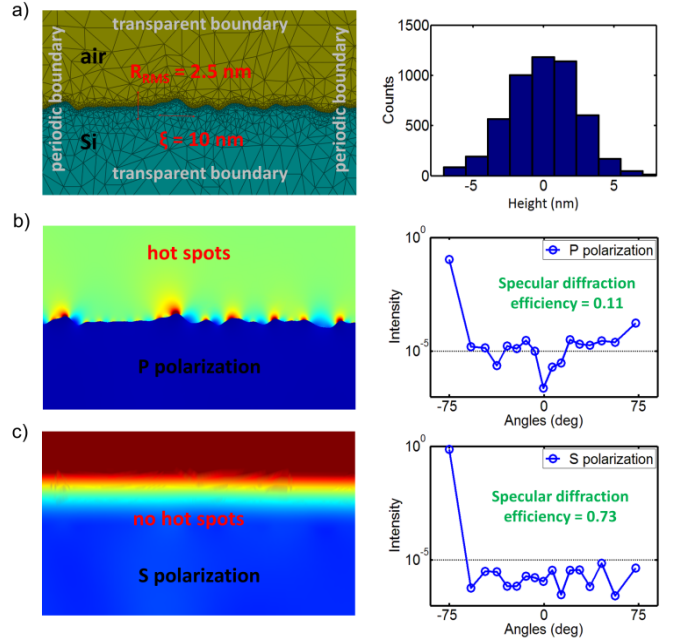

Figure 1: Scattering simulation of one of the generated surface roughness for a plane wave incident at 75° at a wavelength of 600 nm. (a) Local grid structure after one refinement step (left) and the Gaussian distribution of surface heights (right). Near field intensity image and far field intensity angular distribution for (b) *P* polarization and for (c) *S* polarization (with -75° meaning the specular reflection).

JCMSuite permits adaptive mesh refinement, i.e., after a pre-generated grid (following the curvature of the geometry), local grid refinements are applied as a function of the previously solved field amplitude gradients and a new refined mesh is calculated. These steps can be iterated to achieve adequate convergence and necessary precision. Faster convergences can be achieved when using higher FEM degrees. In our simulations, computational costs and ellipsometric angle convergences as a function of the refinement steps and the FEM degree were also investigated. For the current morphologies ($\xi$ and $R_{RMS}$ are smaller than *λ*), a suitable compromise for computation costs was 1 refinement step and a FEM degree of 3, with which a *Ψ* convergence smaller than $10^{-2}$ and a *Δ* convergence smaller than $10^{-1}$ degree were achieved.

The FEM simulated spectra were fitted with a planar thin layer structure using the transfer matrix method [11], where the surface roughness is considered to be an effective medium volumetrically composed of 50-50% of the two media [10, pp. 181-184]. D. E. Aspnes et al. concluded that the Bruggeman EMA showed the best fit results for the ellipsometric evaluations of various rough surfaces [15], and has been extensively used for such evaluations since then. The simplest single layer EMA representing the surface roughness (see inset in Fig.2a) has only one fit parameter, namely its thickness value ($d_{EMA}$). The void is kept fixed at 50% as mentioned above, as the screening parameter as well, kept fixed at a value of 1/3, representing spherical inclusions in the EMA model. The fitting algorithm minimizes the mean square error (*MSE*), indicating the merit of fit. In our case, for one fitted parameter,

$$MSE = \sqrt{\frac{1}{N-2}\sum_{j=1}^{N}\left\{\left(\Psi_j^{FEM} - \Psi_j^{EMA}\right)^2 + \left(\Delta_j^{FEM} - \Delta_j^{EMA}\right)^2\right\}},$$

where the superscripts '*FEM*' and '*EMA*' of *Ψ* and *Δ* indicate the FEM simulation values and the fitted EMA values respectively, while *N* is the number of independently simulated spectral points.

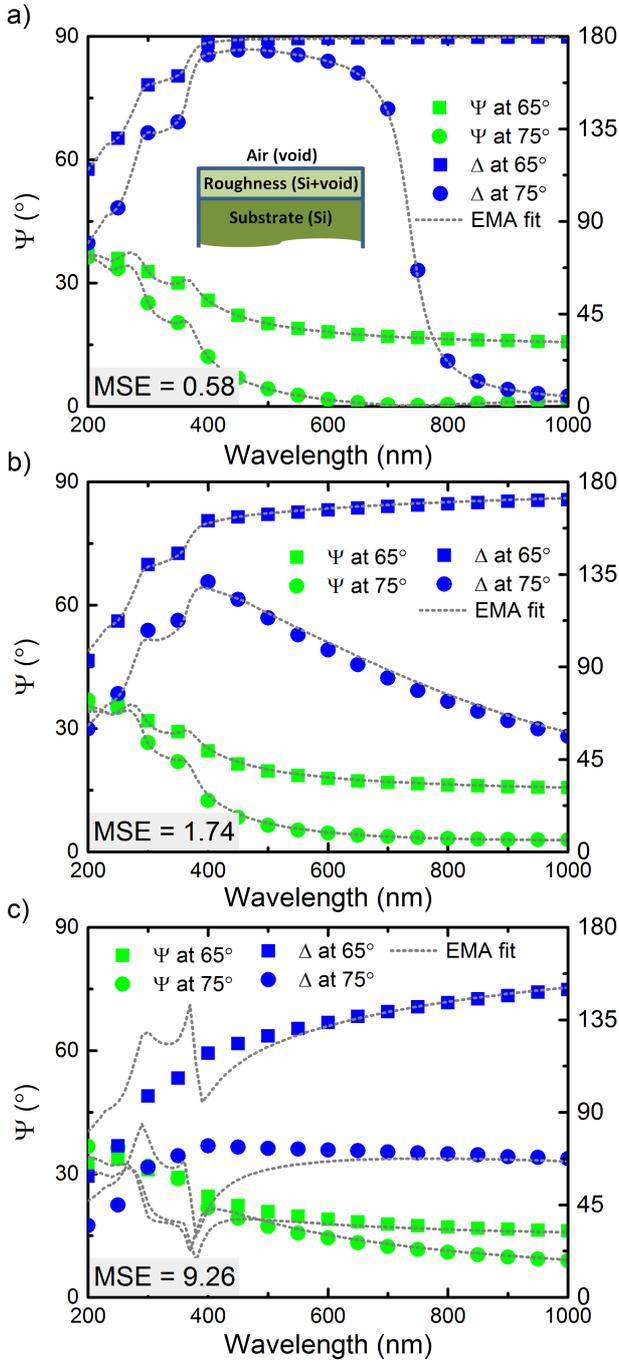

Figure 2: FEM simulations of $\Psi$ and $\Delta$ spectra fitted with an EMA surface roughness (see the model in the inset) of the samples with a nominal correlation length ($\xi$) of 10 nm and with a nominal root mean square roughness of (a) 1 nm, (b) 5 nm and (c) 10 nm. Mean square errors (*MSE*) are also included in the graphs.

### 3. Results and discussion

The small surface features cause high intensity spots in near field around the sharp features of surface protrusions for the *P* polarization, which are not present for the *S* polarization (see left images of Figs. 1b and 1c, respectively, for plane wave illumination at an angle of incidence of 75°, and a wavelength of 600 nm). The difference for the two polarizations is clearly accountable in the diffracted far field intensity values as well. The right-hand side images of Figs. 1b and 1c show the angular intensity distributions of the two polarizations. Although the ellipsometric angles were calculated solely from the 0[th] order (specular) diffracted amplitudes, it is interesting to note that apart from the specular intensity differences (diffraction efficiency of 0.11 for *P* polarization and 0.73 for *S* polarization), there is (generally) an order of magnitude difference in the higher order diffracted angles between the two polarizations. For cases where $\lambda$ is much larger than the typical feature size of the surface roughness, non-specular scattering would be negligible for ellipsometric considerations and also EMA models are applicable. At wavelengths comparable to the typical feature size, scattering starts to dominate and EMA clearly fails to describe the roughness. To demonstrate this phenomenon, Fig. 2 shows the EMA-fitted spectra on simulations with an increasing $R_{RMS}$ value ($R_{RMS}$ = 1, 5 and 10 nm for Figs. 2a, 2b and 2c, respectively) for an identical $\xi$ = 10 nm. For the $R_{RMS}$ = 1 nm ($d_{EMA}$ = 0.3 nm), an almost perfect match can be fitted, while for the $R_{RMS}$ = 5 nm case ($d_{EMA}$ = 6.7 nm), small deviations at the UV part of the spectra start to appear with an increase in the *MSE* value. Finally, for the $R_{RMS}$ = 10 nm case ($d_{EMA}$ = 24 nm), fitting on the whole spectra would be inappropriate, biasing the evaluated roughness; the fit shown in Fig. 2c was made in a wavelength range of 800 - 1000 nm only (*MSE* = 1), and the $\Psi$ and $\Delta$ angles were generated (extrapolated) to the whole range to point out the huge deviations from the FEM simulations below $\lambda$ = 600 nm.

For the following discussion, only the simulations where the whole spectral range can be fitted with the EMA model (*MSE* < 3) are considered. Fig. 3 summarizes the dependence of $d_{EMA}$ on the $R_{RMS}$ and $\xi$ values. The most conspicuous effect is that separate relations can be established between the $d_{EMA}$ and the $R_{RMS}$, depending on $\xi$. Interestingly, quadratic relation fits are much more accurate than simple linear ones at these parameter ranges. Additionally, the different "curvatures" indicate that ellipsometry is more sensitive to sharper surface roughness features in the microscopic regime, i.e., for shorter $\xi$ values, fitted dEMA increases at a higher pace as a function of RRMS than for longer $\xi$ values. This effect agrees well with the conclusion made in Ref. [8] that ellipsometry is sensitive on roughness only on relatively short length scales, also demonstrated by 2 linear fits with different slopes in Ref. [2]. In other words, the high-wavenumber contributions of the power spectral density of the surface points dominate the polarization change.

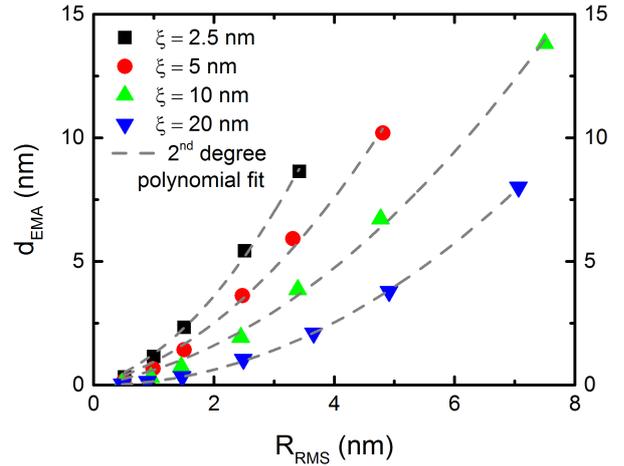

Figure 3: 2$^{nd}$ degree polynomial correlation between the root mean square roughness ($R_{RMS}$) and the thickness of the effective medium roughness ($d_{EMA}$) for different correlation lengths ($\xi$).

The quadratic relation between $d_{EMA}$ and $R_{RMS}$ was also shown to exist in Ref. [16], where the change in polarization due to the interaction of light with the microscopically rough surface was calculated by second-order Rayleigh-Rice formalism (developed by Franta and Ohlídal [17]) and fitted to the EMA calculations. Furthermore, Yanguas-Gil et al. [18] calculated a small correlation length approximation of the Rayleigh-Rice theory for self-affine surfaces. Such surfaces have $R_{RMS}$ values that scale as $L^\alpha$, were $\alpha$ is the roughness exponent, an additional characteristic parameter originating from the dynamics of roughness growth. In the calculations, a $d_{EMA} \sim R_{RMS}^2/\xi^\alpha$ relationship was proven. Similarly to the interpretation done in Ref. [18], that the average surface slope ($R_{dq}$, root mean square average of the local slope, see Ref. [19]) scales as $R_{RMS}/\xi^\alpha$, the $d_{EMA}$ value can be plotted as a function of the product of this $R_{dq}$ and the $R_{RMS}$ value. Fig. 4 reveals a linear correlation for the present study. Excellent linear fit is achieved for $R_{RMS}*R_{dq}$ values smaller than 2 nm. For larger values, downward deviations from the extrapolated line appear, hinting at higher order corrections in the Rayleigh-Rice formalism with, for example, a secondary effect of $\xi$ on $d_{EMA}$ at a unique $R_{RMS}*R_{dq}$ value (see inset in Fig. 4). The linear relationship, mentioned in the many experimental reports [1-7], between $R_{RMS}$ measured by AFM and $d_{EMA}$ measured by ellipsometry can be explained by the fact that the slopes remain constant in most roughening dynamics [1].

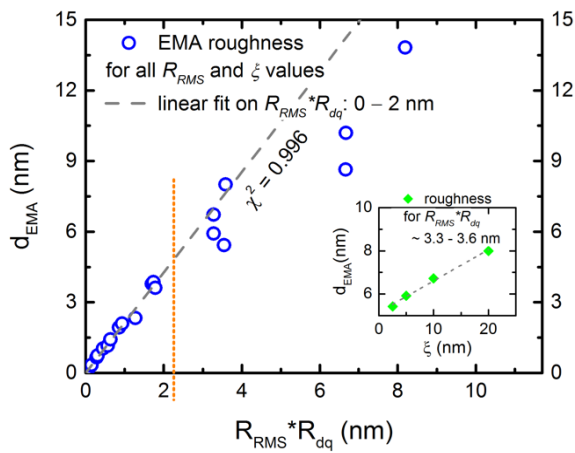

Figure 4: Correlation between the product of RMS roughness and RMS slope ($R_{RMS}*R_{dq}$) and the thickness of the effective medium roughness ($d_{EMA}$) with linear fit for abscissa values smaller than 2 nm. The inset shows the secondary effect of correlation length ($\xi$) on $d_{EMA}$ for points which have an $R_{RMS}*R_{dq}$ value of ~3.4 nm.

## 4. Conclusions

Finite-element method proves to be a very useful tool to simulate the ellipsometric response of light reflected from microscopic stochastic surface roughness. Not hindered by the sample preparation and the experimental conditions, one can define ideal Gaussian random surfaces with well-defined morphological parameters, such as the RMS roughness and the correlation length in our case. As the effective medium approximation is the most widely used model in ellipsometric evaluations of surface roughness, the present paper focused on the correlation between the fitted EMA thickness and the RMS roughness. A linear relationship between the $d_{EMA}$ and the product of the RMS roughness and the average surface slope has been found for smaller $d_{EMA}$ values, in accordance with the results analytically calculated with Rayleigh-Rice formalism and with the vast experimental measurements reported in previous papers. The deviation from the linear relationship foreshadows further corrections between the relationship of $d_{EMA}$ and the surface morphological parameters.

**Acknowledgments**

This work was supported by the OTKA grant Nrs. K81842 K115852, by ENIAC E450EDL project, the Hungarian–French Intergovernmental S&T Cooperation Programme TÉT as well as the M-ERA-NET project Nr. 117847.